\documentstyle[12pt]{article}
\normalbaselines

\begin{document}
\renewcommand{\thefootnote}{\fnsymbol{footnote}}

\ \\

\vspace{-2.cm}

\begin{flushright}

{\raggedleft SWAT/95/72\\
hep-lat/9505001\\[1.cm]}

\end{flushright}

\begin{center}
{\LARGE\baselineskip0.9cm
Some approximate analytical methods in

the study of the self-avoiding loop model

with variable bending rigidity and the

critical behaviour of the strong

coupling lattice Schwinger model

with Wilson fermions\\[1.cm]}
{\large K. Scharnhorst\footnote[2]{E-mail:
k.scharnhorst @ swansea.ac.uk}
}\\[0.3cm]
{\small University of Wales, Swansea

Department of Physics

Singleton Park

Swansea, SA2 8PP, U.K.}\\[1.5cm]
\end{center}
\renewcommand{\thefootnote}{\arabic{footnote}}

\thispagestyle{empty}

\begin {abstract}
Some time ago Salmhofer demonstrated the equivalence of the
strong coupling lattice Schwinger model with Wilson fermions
to a certain 8-vertex model which can be understood as a
self-avoiding loop model
on the square lattice with bending rigidity $\eta = 1/2$ and monomer
weight $z = (2\kappa)^{-2}$. The present paper applies two
approximate analytical methods to the investigation of critical
properties of the self-avoiding loop model with variable bending
rigidity, discusses their validity and makes comparison with
known MC results. One method is based on the independent loop
approximation used in the literature for studying phase
transitions in polymers, liquid helium and cosmic strings.
The second method relies on the known exact solution of the
self-avoiding loop model with bending rigidity $\eta = 1/\sqrt{2}$.
The present investigation confirms recent findings that
the strong coupling lattice Schwinger model becomes critical
for $\kappa_{cr} \simeq 0.38-0.39$. The phase transition
is of second order and lies in the Ising model universality
class. Finally, the central charge of the strong coupling
Schwinger model at criticality is discussed and predicted
to be $c = 1/2$.\\
\end{abstract}

\newpage

\section{Introduction}

Recently, the strong coupling lattice Schwinger model with
Wilson fermions ($N_f = 1$) has received some attention
\cite{gaus1}--\cite{karsch} following
work by Salmhofer \cite{salm} who has shown that it
is equivalent to a certain 8-vertex model (a 7-vertex model, more
precisely) which also can be
understood as a self-avoiding loop model
on the square lattice with bending rigidity $\eta = 1/2$ and monomer
weight $z = (2\kappa)^{-2}$. Beyond its toy character
interest in the lattice Schwinger model (${\rm QED_2}$) mainly
derives from the similarity of some of its major features with
those of QCD in 4D. However, because the result of Salmhofer \cite{salm}
is related to the polymer (hopping parameter) expansion
of the fermion determinant \cite{mont1}, \cite{kar1} the
strong coupling Schwinger model is also interesting from
the point of view of the dynamical fermion problem within
lattice gauge theory. While some investigations have been devoted
to the polymer expansion of the fermion determinant in the
case of staggered fermions \cite{burk}--\cite{mut} to date
almost no attention has been paid to the corresponding
case of Wilson fermions \cite{luo} due to
the additional difficulties involved in general. However,
while in the strong coupling limit the system with staggered
fermions (QCD, QED) reduces to a pure monomer-dimer system \cite{ros} the
same is not true for Wilson fermions as the investigation of
Salmhofer \cite{salm} demonstrates. The equivalence of the
strong coupling lattice Schwinger model with Wilson
fermions to a self-avoiding loop model enables certain methods
used in other branches of physics, e.g. in condensed matter physics (polymers,
defect mediated phase transitions) and in cosmic string physics,
to be exploited in its investigation \cite{wie1}--\cite{cope5}.
At the same time, its equivalence
(in another language) to some 8-vertex model \cite{bax} makes
further results available.\\

Self-avoiding loop models \cite{genn}, \cite{rys1}
have a long history due to their
prominent role in polymer physics as well as
their attraction as a simple problem of non-Markovian
nature. In addition, systems of closed non-crossing lines
or systems which can be approximated by them appear in
a variety of contexts ranging from condensed matter physics
through cosmology to quantum field theory which generates
common interest for appropriate model building \cite{klein}, \cite{wie1}.
Recently, quantum field theoretic methods have
been exploited to study the critical behaviour of
self-avoiding loop models in two dimensions
\cite{card1}--\cite{dup1}. Somewhat lesser attention has been paid
so far to the self-avoiding loop model with variable bending
rigidity (while for open chains with bending rigidity
a number of investigations exists, e.g.,
\cite{moon} and references therein).
Beyond the work of M\"user and Rys \cite{mus1}
certain insight in this direction has been obtained in connection
with the study of 2D vesicles \cite{leib}--\cite{cam}.\\

 From the point of view of the 8-vertex model, a general solution
to the self-avoiding loop model with variable bending rigidity
on the square lattice is not known. However, for the special
case $\eta = 1/\sqrt{2}$ the free-fermion condition \cite{fan1},
\cite{fan2} is fulfilled and it can be solved exactly
\cite{pri1}--\cite{blu}. This way one point on the critical line
of the the self-avoiding loop model with variable bending
rigidity is known exactly and consequently one may use
methods of perturbative nature in its neighbourhood to
approximately find the critical line there by analytical methods.\\

The plan of the paper is as follows. In section 2 we shortly
review the relevant facts concerning the lattice Schwinger
model with Wilson fermions and discuss the relation of recent
exact studies of its partition function on finite lattices
\cite{gaus3}, \cite{karsch} to the
earlier MC results of M\"user and Rys \cite{mus1}.
Section 3 is devoted to the approximate analytical
study of the self-avoiding loop
model (SALM) with variable bending rigidity by means of the
independent loop approximation. Section 4 then explores
the application of the exact solution of the SALM
with bending rigidity $\eta  = 1/\sqrt{2}$ to the
study of the critical behaviour of the SALM
in a neigbourhood of it in the
relevant parameter space. Section 5 finally discusses the
picture emerging from the present investigation paying
special attention to the
central charge of the SALM with
variable bending rigidity on the critical line.\\

\section{The strong coupling Schwinger model with \hfill\break
Wilson fermions}

The partition function $Z_\Lambda$ of the Schwinger model with
Wilson fermions (with Wilson parameter $r = 1$)
on a certain lattice $\Lambda$
is given by the standard expression

\parindent0.em

\begin{eqnarray}
\label{B1}
Z_\Lambda &=& \int DU D\psi D\bar\psi\ \ {\rm e}^{-S}\ \ \ ,
\end{eqnarray}

where $D$ denotes the multiple integration on the lattice. The
action $S$ is defined by

\begin{eqnarray}
\label{B2}
S &=& S_F + \beta S_G\ \ \ \ \ ,\\
\vspace{0.3cm}\nonumber\\
\label{B3}
S_F &=& \sum_{x\in \Lambda} \left( {1\over 2}
\sum_\mu \left(\bar\psi(x+\hat{\mu})(1+\gamma_\mu)
U_\mu^\dagger(x) \psi(x)\right.\right.\nonumber\\
\vspace{0.3cm}\nonumber\\
&&\ \ \ +
\left.
\bar\psi(x)(1-\gamma_\mu) U_\mu^\dagger(x) \psi(x+\hat{\mu})
\right) - M \bar\psi(x)\psi(x)\Bigg)\ \ \ \ \ ,
\end{eqnarray}

and $U_\mu =\exp{[-i A_\mu]}$, $M = 2 + m$, $\beta = 1/g^2$.
$S_G$ is the standard Wilson action and the hopping parameter
$\kappa$ is given by $\kappa = 1/2M$. Salmhofer has shown \cite{salm}
that in the strong (infinite) coupling limit $\beta = 0$ the
partition function $Z_\Lambda$ equals that of an 8-vertex model
(a 7-vertex model due to eq.\ (\ref{B4b}), more precisely)
\cite{bax} with weights (cf.\ fig.\ 1)

\begin{eqnarray}
\label{B4a}
w_1 &=& z\ =\ {1\over 4 \kappa^2}\ =\ M^2\ \ \ \ ,\\
\vspace{0.2cm}\nonumber\\
\label{B4b}
w_2 &=& 0\ \ \ \ ,\\
\vspace{0.2cm}\nonumber\\
\label{B4c}
w_3 &=& w_4\ =\ 1\ \ \ \ ,\\
\vspace{0.2cm}\nonumber\\
\label{B4d}
w_5&=& w_6\ =\ w_7\ =\ w_8\ =\ \eta\ =\ {1\over 2}\ \ \ \ .
\end{eqnarray}

Consequently, one can write

\begin{eqnarray}
\label{B5}
Z_\Lambda &=& Z_\Lambda\left[z,{1\over 2}\right]\ \ \ \ ,\\
\vspace{0.3cm}\nonumber\\
\label{B6}
Z_\Lambda[z,\eta] &=&
\sum_L\ z^{\vert\Lambda\vert - \vert L\vert}\ \eta^{C(L)}\ \ \ ,
\end{eqnarray}

where $L$ denotes any self-avoiding loop configuration,
$\vert L\vert$ and $C(L)$ are the number of links and corners respectively
a polymer configuration $L$ is built of,
and $\vert\Lambda\vert$ is the number of lattice points of the
lattice $\Lambda$. $Z_\Lambda[z,\eta]$ is the partition function
of a self-avoiding loop model (SALM) with monomer weight $z$ and bending
rigidity $\eta$. The same expression can of course also be
obtained for non-compact $\rm QED_2$.\\

\parindent1.5em

 From the point of view of lattice field
theory it is interesting to know the phase structure of the
lattice Schwinger model. For free fermions ($\beta = \infty$)
the critical value of the hopping parameter reads
$\kappa_{cr}(\beta = \infty) = 1/4$. In order to pin down the
critical line for $\beta < \infty$ it is of particular
interest to know where it ends ($\beta = 0$). There is
a critical point for $\kappa_{cr}(\beta = 0) = \infty$ because then the
strong coupling Schwinger model reduces to a 6-vertex model
whose behaviour is known from its exact solution \cite{salm},
\cite{bax}. This point however is believed to be isolated
and not to be the end point of the critical line starting at
$\kappa_{cr}(\infty) = 1/4$ \cite{gaus2}.
Recently, exact studies of the partition function of the
strong coupling Schwinger model have been made on finite lattices
\cite{gaus3}, \cite{karsch}. It has been found
$\kappa_{cr}(0) \simeq 0.38-0.39$ and that the phase transition
is likely a continuous one (second order or higher) \cite{karsch}.\\

It is worthwhile to compare the result obtained in
\cite{gaus3}, \cite{karsch} with the MC investigation of the SALM
with variable bending rigidity undertaken by
M\"user and Rys \cite{mus1} (see also \cite{mus2} for
some computational background). Their investigation
has been inspired by the generalized loop model of Rys and
Helfrich \cite{rys1}.
M\"user and Rys \cite{mus1} employ a different parameter
set than $\{z,\eta\}$ which we are going to describe first.
Their language is thermodynamic in spirit and their
parameters temperature and line stiffness $\{T,s\}$ are
introduced the following way

\parindent0.em

\begin{eqnarray}
\label{B7}
z &=& {\rm e}^{\displaystyle\ (1-s)/T}\ \ \ \ ,\\
\vspace{0.3cm}\nonumber\\
\label{B8}
\eta &=& {\rm e}^{\displaystyle -s/T}
\end{eqnarray}

which in turn entails

\begin{eqnarray}
\label{B9}
T &=& {1\over \ln{\displaystyle{z\over\eta}}}\ \ \ \ \ ,\\
\vspace{0.3cm}\nonumber\\
\label{B10}
s &=& {\ln\eta\over\ln{\displaystyle{\eta\over z}}}\ \ \ \ \ .
\end{eqnarray}

For positive temperatures $T$ negative values of the line stiffness
$s$ correspond to values of the bending rigidity $\eta > 1$
(i.e., bending preferred) and positive values of $s$ to
$\eta < 1$ (i.e., bending is costly).
The Jacobian ${\cal F}$ of this coordinate transformation
 from $\{z,\eta\} \in \{[0,\infty),[0,\infty)\}$ to
$\{T,s\} \in \{[0,\infty),(-\infty,\infty)\}$ reads

\begin{eqnarray}
\label{B12}
{\cal F} &=& z \eta \left[ \ln{z\over\eta}\right]^3\\
\vspace{0.3cm}\nonumber\\
\label{B13}
&=& {1\over T^3}\ \ {\rm e}^{\displaystyle\ (1-s)/T}
\end{eqnarray}

and has consequently singular lines beyond
the given range of the map.
Figure 2 displays the result of
M\"user and Rys \cite{mus1} (adapted from their fig.\ 2)
for the critical line of the loop model with line stiffness
and fig.\ 3 displays the same information in $\{z,\eta\}$
coordinates (for further comments see the figure captions).
In regions II and III the system at criticality is found to exhibit
Ising-like behaviour while in region I some non-universal behaviour
is seen. For better orientation the ordinary loop gas ($\eta = 1$)
result is specially shown in figs.\ 2, 3 \cite{kar2}--\cite{kar3}.
One immediately recognizes that the results found for the
strong coupling Schwinger model \cite{gaus3}, \cite{karsch}
fits well onto the critical line given by M\"user and Rys.
Moreover, the MC result of M\"user and Rys also well
agrees with the exactly known critical point for the free
fermion model ($\eta = 1/\sqrt{2}$) to be discussed in section 4.
For the ordinary loop gas it has also been found numerically
\cite{kar3} that critical exponents (and amplitudes, in part) agree well
with those of the Ising model. The free fermion model
($\eta = 1/\sqrt{2}$, see section 4) of course also lies
in the Ising universality class. This immediately suggest
(for a further discussion see section 5) that in general the
SALM with variable bending rigidity at criticality lies in the
Ising universality class (in the parameter regions II, III).
 From this we immediately infer that also the strong coupling
lattice Schwinger model ($\eta = 1/2$) belongs to this class.
In \cite{karsch} however a critical exponent $\nu \simeq 0.63$
has been reported for the strong coupling Schwinger model
which is quite off the Ising result $\nu = 1$. The discrepancy
very likely stems from finite size effects of the small
lattices on non-quadratic domains investigated. These non-quadratic
domain lattices however can be exploited in other ways
as we will see in section 5.\\

\section{The independent loop approximation}

Inasmuch as exact expressions for the partition function
(\ref{B6}) for general $\eta$ are not available analytical
attempts to understand the phase structure of the self-avoiding
loop model with variable bending rigidity have to rely
on certain approximations. A method also applied in related
situations in condensed matter physics and cosmic string
physics is the so-called 'independent loop approximation'
\cite{wie1}-\cite{cope5}, \cite{wie2}, \cite{weg}. The approximation
is approached by writing the partition function $Z_\Lambda[z,\eta]$
as a sum over partition functions with a fixed number $l$ of
(polymer) loops.

\begin{eqnarray}
\label{C1}
Z_\Lambda[z,\eta] &=&
z^{\vert\Lambda\vert}\ \ \sum_{l=0}^\infty\ Z_\Lambda[l,z,\eta]
\end{eqnarray}

The approximation now made is to express the $l$-loop partition
function $Z_\Lambda[l,z,\eta]$ exclusively by means of the single
loop partition function $Z_\Lambda[1,z,\eta]$

\begin{eqnarray}
\label{C2}
Z_\Lambda[l,z,\eta] &=& {1\over l!}\
\left[\ Z_\Lambda[1,z,\eta]\ \right]^l  \ \ \ \ ,
\end{eqnarray}

leading to

\begin{eqnarray}
\label{C3}
Z_\Lambda[z,\eta] &=&
z^{\vert\Lambda\vert}\ \ {\rm e}^{\displaystyle\ Z_\Lambda[1,z,\eta]}\ \ \ \ .
\end{eqnarray}

This approximation can be expected to give reasonable
results for those parameter regions where the loop system
is sufficiently dilute (filling in the average only a certain
fraction of the lattice $\Lambda$). According to eq.\ (\ref{C3})
the investigation now may concentrate on the single loop
partition function $Z_\Lambda[1,z,\eta]$. One can easily convince
oneself that in the independent loop approximation the average
number of loops in the system is given by the value of the
single loop partition function \cite{cope1}, eq.\ (56).
The free energy density $f$ reads in the independent loop
approximation ($\beta_T = 1/T$)

\begin{eqnarray}
\label{C4}
\beta_T f(z,\eta)&=& - \lim_{\vert\Lambda\vert\longrightarrow\infty}\
{1\over\vert\Lambda\vert}\
\ln Z_\Lambda[z,\eta]\ = \ -\
\ln z\ -\ \lim_{\vert\Lambda\vert\longrightarrow\infty}\
{Z_\Lambda[1,z,\eta]\over\vert\Lambda\vert}\ \ \ .
\end{eqnarray}

\parindent1.5em

To proceed further, the single loop partition function can now be written
as sum over the loop length

\parindent0.em
\begin{eqnarray}
\label{C5}
Z_\Lambda[1,z,\eta] &=&
\sum_{k=1}^\infty\ z^{-2k}\ {\cal Z}_\Lambda[2k,\eta]
\end{eqnarray}

where ${\cal Z}_\Lambda[2k,\eta]$ is the conformational
partition function of a single loop of length $2k$ (Here we already have
taken into account that on a square lattice the length
of a loop is always even.) \cite{cope4}. The conformational partition
function is represented then as sum over all single loop
configurations $L$ with length $2k$

\begin{eqnarray}
\label{C6}
{\cal Z}_\Lambda[2k,\eta] &=&
\sum_{L,\ \vert L\vert = 2k}\ \eta^{C(L)}\ =\
\sum_{C=0}^{2k}\ N(2k,C)\ \eta^C\ \ \ \ .
\end{eqnarray}

\parindent1.5em

Let us start with the consideration of the ordinary loop
model ($\eta = 1$). In this case ${\cal Z}_\Lambda[2k,1]$
denotes the total number of possibilities to place a
self-avoiding loop with length $2k$ on the lattice $\Lambda$.
It can be expressed by means of the number $p_{2k}$ of
$2k$-step self-avoiding loops per lattice site
which is a standard quantity that has been investigated
in the literature.

\parindent0.em

\begin{eqnarray}
\label{C7}
p_{2k}&=&\ \lim_{\vert\Lambda\vert\longrightarrow\infty}\
{{\cal Z}_\Lambda[2k,1]\over\vert\Lambda\vert}
\end{eqnarray}

The $n\rightarrow 0$ limit of the lattice $O(n)$ spin model provides us
now just with the information necessary to study the critical
behaviour \cite{genn} (see also, e.g., \cite{card1}, sect.\ 2).
For large $k$ $p_{2k}$ reads \cite{card2}

\begin{eqnarray}
\label{C8}
p_{2k}&\stackrel{k\rightarrow\infty}{=}&
B\ \mu^{2k}\ [2k]^{-2\nu-1}\ +\  ... \ \ \ .
\end{eqnarray}

Here $\mu$ denotes the connective constant
(effective coordination number)
for the self-avoiding walk problem on the given lattice
$\Lambda$ \cite{hamm} and B is some lattice dependent
constant. The (universal) critical exponent $\nu$ is believed
to be given in two dimensions by $\nu = {3\over 4}$
(obtained on a honeycomb lattice)
\cite{nien1}, \cite{nien2}. Inserting (\ref{C8}) into
eq.\ (\ref{C5}) one finds

\begin{eqnarray}
\label{C9}
Z_\Lambda[1,z,1] &=&
\vert\Lambda\vert\ B\ \sum_{k=1}^\infty\ \
[2k]^{-5/2}\ \left({\mu\over z}\right)^{2k}\ \ \ \ .
\end{eqnarray}

This is a justified approximation because we are
mainly interested in the critical domain which
is related to the $k\rightarrow\infty$ behaviour.
 From eq.\ (\ref{C9}) one easily recognizes that the critical point is given
by $z_{cr} = \mu$. Most recent (precise) estimates
for $\mu$ on the square lattice can be found in
\cite{gut1}--\cite{con}. We keep here only a few digits and
write $z_{cr} = \mu = 2.638$ ($T_{cr} =1.031$). Eq.\ (\ref{C9}) inserted into
eq.\ (\ref{C4}) gives immediately the free energy and
one recognizes that the phase transition at
$z_{cr} = \mu = 2.638$ found within the independent
loop approximation is of second order.
Using (\cite{lind}, \cite{true})

\begin{eqnarray}
\label{C10}
F(x,k)&=&\sum_{n=1}^{\infty}\ {x^n\over n^k}
\ =\ \Gamma(1-k)\ (-\ln x)^{k-1}\ +\ \sum_{n=0}^{\infty}\
\zeta(k-n)\  {(\ln x)^n\over n!}
\end{eqnarray}

one reobtains for the critical exponent of the
specific heat $\alpha$ the hyperscaling relation
$\alpha = 2-2\nu$ entailing in the independent loop
approximation $\alpha = 1/2$ which is to be confronted with
the expected Ising result $\alpha = 0$ \cite{kar3}.\\

\parindent1.5em

We are now prepared to study the general case with
variable bending rigidity $\eta$. First we have to
find an appropriate expression for the number $N(2k,C)$
of self-avoiding loops with length $2k$ and $C$ corners.
Let us count first the number of random non-backtracking
walks of length $2k$ with $C$ corners \cite{cope4}.
It should be stressed that the following argument
does not depend on the dimension of the lattice. There
are $2k-1$ vertices available the $C$ corners can be
placed at, i.e., there are ${2k-1\choose C}$
possibilities to do so. To each prospective corner exist
$h= (2d -1) -1$ ways of bending where $d$ is the dimension of
a (hyper)cubic lattice (in our case of a square lattice $d=2$).
$(2d -1)$ is here the non-backtracking dimension of the lattice
and it has to be diminished by $1$ corresponding to
the straight line choice. Consequently, we find

\parindent0.em

\begin{eqnarray}
\label{C12}
N_{NB}(2k,C)&=& \vert\Lambda\vert\ {2k-1\choose C}\
h^C\ \ \ \ .
\end{eqnarray}

Using eq.\ (\ref{C6}) the corresponding (non-backtracking) conformational
partition function reads then

\begin{eqnarray}
\label{C13}
{\cal Z}_\Lambda[2k,\eta]_{NB} &=& \vert\Lambda\vert\
\left[\ 1+h\eta\ \right]^{2k-1}\ \ \ \ .
\end{eqnarray}

We now obtain an approximation to the $2k$-step self-avoiding walk
(SAW) conformational partition
function by simply replacing $h= (2d -1)[1 -1/(2d -1)]$
by $h= \bar\mu [1 -1/(2d -1)]$ (this is based on the assumption
that the self-avoidance constraints effectively encoded in $\bar\mu$
are independent of whether propagation is straight or bent)
where $\bar\mu$ is a certain effective coordination number
to be determined in a moment.
This yields in our case

\begin{eqnarray}
\label{C14}
{\cal Z}_\Lambda[2k,\eta]_{SAW} &=& \vert\Lambda\vert\
U(2k)\ \left[\ 1+{2\bar\mu\eta\over 3}\ \right]^{2k-1}\ \ \ \ .
\end{eqnarray}

The additional factor $U(2k)$ also to be determined below
takes care of some additional length dependence which might
show up in the transition from non-back\-track\-ing to SAW.
Specializing eq.\ (\ref{C14}) to $\eta = 1$ we find the total number
of SAW of length $2k$ on the square lattice which has to
be confronted with the standard expectation \cite{card2} for
large $k$

\begin{eqnarray}
\label{C15}
c_{2k}&=&\ \lim_{\vert\Lambda\vert\longrightarrow\infty}\
{{\cal Z}_\Lambda[2k,1]_{SAW}\over\vert\Lambda\vert}\
\stackrel{k\rightarrow\infty}{=}\
A\ \mu^{2k}\ [2k]^{\gamma -1}\ +\ \ ...
\end{eqnarray}

where $A$ is some lattice dependent constant and
$\gamma = 43/32$ \cite{nien1}, \cite{nien2}. From
eq.\ (\ref{C15}) we immediately find

\begin{eqnarray}
\label{C15a}
\bar\mu&=& {3\over2}\ (\mu-1)\ \ \ \ ,\\
\vspace{0.3cm}\nonumber\\
\label{C15b}
U(2k)&=& A\ \mu\ [2k]^{\gamma -1}
\end{eqnarray}

\parindent1.5em

Now, we need to know the conformational partition
function for the self-avoiding loop problem.
In order to be able to make further progress let us assume that
$N(2k,C)$ for self-avoiding loops is just a certain
fraction $M(2k)$ of $N_{SAW}(2k,C)$ at least for large $k$
independent of the number of corners $C$. According to eq.\ (\ref{C6})
then we can write

\parindent0.em

\begin{eqnarray}
\label{C16a}
{\cal Z}_\Lambda[2k,\eta] &=& M(2k)\ {\cal Z}_\Lambda[2k,\eta]_{SAW}
\end{eqnarray}

which reads after having taken into account eqs.\ (\ref{C14}),
(\ref{C15a}), (\ref{C15b})

\begin{eqnarray}
\label{C16b}
{\cal Z}_\Lambda[2k,\eta] &=& \vert\Lambda\vert\ M(2k)\
A\ \mu
\left[\ 1+(\mu-1)\eta\ \right]^{2k-1}\ [2k]^{\gamma -1}\ \ \ \ .
\end{eqnarray}

$M(2k)$ is the fraction to be determined. We here simply
ignore the fact that for any loop the number of corners
is even, necessarily. This is justified for the
study of the $k\rightarrow\infty$ behaviour we are primarily
interested in. For $\eta = 1$
we already have displayed an expression (eq.\ (\ref{C8}))
which now serves as reference expression to determine $M(2k)$.
We obtain

\begin{eqnarray}
\label{C17}
M(2k)&=& {B\over A}\ [2k]^{-2\nu -\gamma}
\end{eqnarray}

leading to

\begin{eqnarray}
\label{C18}
Z_\Lambda[1,z,\eta] &=&
\vert\Lambda\vert\ {B\ \mu\over\left[\ 1+(\mu-1)\eta\ \right]}\
\sum_{k=1}^\infty\ \ [2k]^{-5/2}\
\left({ 1+(\mu-1)\eta\over z}\right)^{2k}\ \ \ \ .
\end{eqnarray}

Consequently, the critical line is found to be

\begin{eqnarray}
\label{C19}
\eta_{cr}(z_{cr}) &=& {(z_{cr}-1)\over(\mu-1)}\
\ \ \ \ .
\end{eqnarray}

This translates into the $\{T,s\}$ coordinate system as

\begin{eqnarray}
\label{C20}
s_{cr}(T_{cr})&=& T_{cr}\
\ln\left[\
{{\rm e}^{\displaystyle\ 1/T_{cr}}\over\mu}\ -\ \mu\ +\ 1\ \right]\ \ \ \ .
\end{eqnarray}

It should be emphasized that the result of our approximate
consideration (eq.\ (\ref{C20})) entails $s_{cr}\rightarrow 1$
for $T_{cr}\rightarrow 0$. This is well in line with expectations
spelled out in \cite{mus1}. From the above equations we
obtain for the strong coupling Schwinger model
$z_{cr} = 1.819$ ($T_{cr} = 0.774$, $s_{cr} = 0.537$). We find
for the critical hopping parameter

\begin{eqnarray}
\label{C21}
\kappa_{cr}(0)&=& {1\over \sqrt{2 (\mu + 1) }}\ =\ 0.371\ \ \ \ .
\end{eqnarray}

\parindent1.5em
The critical line (eqs.\ (\ref{C19}), (\ref{C20})) obtained within
the independent loop approximation is plotted in figs.\ 4, 5.
One recognizes that the critical line found analytically agrees
qualitatively quite well with the result of the MC calculation of
M\"user and Rys \cite{mus1}. However, it is clear that the
validity of the independent loop approximation is confined
to the low (polymer) loop density domain. The high density
result of M\"user and Rys \cite{mus1} displayed in region I of
figs.\ 2, 3 cannot be obtained within the present scheme.
It is also well known \cite{wie1}, \cite{wie2} that while
within the independent loop
approximation the critical line can be determined in a
qualitatively correct way results for
critical exponents are correct to a lesser degree.
This also applies to our case as we have seen above.
While we would have expected, e.g., for the critical
exponent $\alpha$ the Ising result ($\alpha = 0$)
within the independent loop approximation we see $\alpha =1/2$.
Finally, it should also be mentioned that the relative
simplicity of the independent loop approximation has its
price because so far no way of its systematic improvement
is known and one consequently has no quantitative control
over the approximation made. Perhaps, this drawback
is offset by the applicability of the approximation
to systems in any dimension.\\

\section{The exact solution of the self-avoiding loop model
with bending rigidity
$ \eta = 1/\surd\overline{2}$ and its use}

While in general the self-avoiding loop model (SALM) with variable bending
rigidity (which is equivalent to a 7-vertex model due to $w_2 = 0$)
cannot be studied exactly so far, there exists an
exact solution to it for $\eta = 1/\sqrt{2}$ first investigated by
Priezzhev \cite{pri1} (see also \cite{pri2}). This solution has
later been rediscovered by Blum and Shapir \cite{blu}
who apparently were unaware of the earlier work of Priezzhev.
The solution relies on the general study of the 8-vertex model by Fan
and Wu \cite{fan1}, \cite{fan2}. They found that the 8-vertex model
is exactly solvable if the free fermion condition

\parindent0.em

\begin{eqnarray}
\label{D1}
w_1 w_2\ +\ w_3 w_4 &=& w_5 w_6\ +\ w_7 w_8
\end{eqnarray}

is fulfilled (cf. fig. 1 for the labelling of the vertices).
Inserting eqs.\ (\ref{B4a})-(\ref{B4d})
into (\ref{D1}) ($\eta$ taken arbitrary here) one immediately finds
that the free fermion condition is fulfilled for $\eta = 1/\sqrt{2}$.
The partition
function for the SALM with bending rigidity
$\eta = 1/\sqrt{2}$ has been found in \cite{pri1}, \cite{blu} by
standard methods. The free energy density $f$ reads

\begin{eqnarray}
\label{D2}
\beta_T f(z,1/\sqrt{2})&=&- {1\over 8\pi^2}\
\int_0^{2\pi} d\theta \int_0^{2\pi} d\phi\
\ln\left[\ 2\ +\ z^2\ +\ 2 z \cos\theta\right.\nonumber\\
\vspace{0.3cm}\nonumber\\
&&\hspace{4.cm}
\left. +\ 2 z \cos\phi\ +\
2 \cos\theta \cos\phi\ \right]\ \ .
\end{eqnarray}

A second order phase transition occurs for $z_{cr} = 2$
($T_{cr} = 2/(3\ln 2) = 0.962$, $s_{cr} = 1/3$) which will be of main
interest to us. There is
of course also a critical point at $z = 0$ related to the
exactly solvable 6-vertex model \cite{bax}, \cite{salm}.
Because the system
can be represented by means of free fermions \cite{sam} the
SALM with bending rigidity $\eta = 1/\sqrt{2}$
lies in the Ising universality class \cite{blu}. In accordance with
this it has been shown (for $w_2$ chosen arbitrarily) that the
partition function of the free fermion model can be expressed in
terms of that of the regular Ising model \cite{bax1}. It finally deserves
mention that the result of the MC calculation of
M\"user and Rys \cite{mus1} is in complete
agreement with the exact solution of the free fermion
model (cf.\ figs. 2, 3).\\

\parindent1.5em

The above exact solution lying on the critical line of the
SALM with variable bending rigidity
is quite useful because this way one may take advantage of
universality arguments to draw conclusions
about the model at criticality
for fairly wide ranges of the bending rigidity $\eta$. This will
be discussed further in section 5. Here we will
study the approximate calculation of the critical line
in a neighbourhood of the model for $\eta = 1/\sqrt{2}$.
This discussion is in a certain sense a generalization of that given in
\cite{pri2}, \cite{pri1}. Let us write the partition function
(\ref{B6}) in the following way.

\parindent0.em

\begin{eqnarray}
\label{D3}
Z_\Lambda[z,\eta] &=&
\sum_{l=0}^{\vert\Lambda\vert}\ z^{\vert\Lambda\vert - 2l}\
\sum_{L,\ \vert L\vert = 2l}\ \eta^{C(L)}\\
\vspace{0.3cm}\nonumber\\
\label{D4}
&=&\sum_{l=0}^{\vert\Lambda\vert}\ z^{\vert\Lambda\vert - 2l}\
\sum_{L,\ \vert L\vert = 2l}\ \sum_{k=0}^{\infty}\
{1\over k!}\ [ C(L) \ln\eta ]^k\\
\vspace{0.3cm}\nonumber\\
\label{D5}
&=&\sum_{l=0}^{\vert\Lambda\vert}\ z^{\vert\Lambda\vert - 2l}\
\ \sum_{k=0}^{\infty}\
{[\ln\eta ]^k\over k!}\  \langle C(L)^k\rangle_{2l}
\end{eqnarray}

Here,

\begin{eqnarray}
\label{D6}
\langle C(L)^k\rangle_{2l}&=&\sum_{L,\ \vert L\vert = 2l}
\ C(L)^k\ \ \ \ .
\end{eqnarray}

One may now express $\langle C(L)\rangle_{2l}$ the following
way.

\begin{eqnarray}
\label{D7a}
\langle C(L)\rangle_{2l}&=&\bar C_{2l}\ N(2l)\\
\vspace{0.3cm}\nonumber\\
\label{D7b}
\bar C_{2l}&=&2l\ n_C(2l)
\end{eqnarray}

$N(2l) = \langle 1\rangle_{2l}$ denotes the
number of (multi) loop
configurations of total length $2l$ on the lattice $\Lambda$
and $n_C(2l)$, $0\leq n_C(2l)\leq 1$,
stands for the average relative density of corners
in the considered loop ensemble of total length $2l$.
The following of course holds for the higher moments
of $C$.

\begin{eqnarray}
\label{D8}
0 \leq \langle C(L)^k\rangle_{2l}&\leq& (2l)^k\ N(2l)
\end{eqnarray}

One can now write

\begin{eqnarray}
\label{D9}
&&\hspace{-1.5cm}Z_\Lambda[z,\eta]\ =\nonumber\\
\vspace{0.3cm}\nonumber\\
&=& z^{\vert\Lambda\vert}\
\sum_{l=0}^{\vert\Lambda\vert}\ N(2l)\ \left[ {\eta^{n_C(2l)}\over z}
\right]^{2l}\
\left[\ 1\ +\ {\langle (C(L)-\bar C_{2l})^2\rangle_{2l}\over 2\ N(2l)}\
(\ln\eta)^2\ ~~+\ ...\ \ \right]
\end{eqnarray}

where $\ ...\ $ stands for a series in higher order correlation functions
of the corner number $C$ and $\ln\eta$. The critical behaviour of
the system is related to large $l$. For $l\rightarrow\infty$
$n_C$ tends to some value $\bar n_C$ and
consequently points $\{z_1,\eta_1\}$,
$\{z_2,\eta_2\}$ on the critical line not to far away from
each other should obey to leading approximation
the equation

\begin{eqnarray}
\label{D10}
{\eta_1^{\bar n_C}\over z_1}&=&{\eta_2^{\bar n_C}\over z_2}\ \ \ \ .
\end{eqnarray}

The contribution of correlation functions of
$C$ should be expected to be of minor importance in
eq.\ (\ref{D9}), leading to corrections to the
leading behaviour only.
Inserting into eq.\ (\ref{D10}) the exactly known critical point
$\{z,\eta\}=\{2,1/\sqrt{2}\}$ of the free fermion model leads in a
neighbourhood of it to the
equation for the critical line

\begin{eqnarray}
\label{D11}
\eta_{cr}(z_{cr})&=&2^{-(\bar n_C+2)/2\bar n_C}\ z_{cr}^{1/\bar n_C}\ \ \ \ .
\end{eqnarray}

The only unknown quantity in this expression is the average
relative corner density $\bar n_C$. Its value
$n_C(2l\rightarrow\infty)$ is related to the
high density polymer limit which is reached for $z\rightarrow 0$.
$\bar n_C$ has been calculated in \cite{pri1}, \cite{pri2}
for the free fermion model and found to have the value $1/2$.
Values for the correlation functions of $C$ for $l\rightarrow\infty$
can also be obtained along the same lines by tedious, but
standard methods (\cite{mon1}, \cite{mon2};
the latter is the English original of ref.\ no.\ 16
in \cite{pri1}). So, we end up with the
following equation for the critical line of the
SALM with variable bending rigidity in the neighbourhood
of the free fermion point (cf.\ also figs.\ 4, 5).

\begin{eqnarray}
\label{D12}
\eta_{cr}(z_{cr})&=&2^{-5/2}\ z_{cr}^2\ \ \ \ .
\end{eqnarray}

This equation reads in $\{T,s\}$ coordinates

\begin{eqnarray}
\label{D13}
T_{cr}(s_{cr})&=& {2\ (2-s_{cr})\over 5\ \ln 2}\ \ \ \ .
\end{eqnarray}

Consequently, we
obtain for the strong coupling Schwinger model
$z_{cr} = 2^{3/4} \approx 1.682$ ($T_{cr} = 4/(7\ln 2) \approx 0.824$,
$s_{cr} = 4/7 \approx 0.571$). This yields
for the critical hopping parameter

\begin{eqnarray}
\label{D14}
\kappa_{cr}(0)&=& 2^{-11/8}\ \approx\ 0.386\ \ \ \ .
\end{eqnarray}

We see (cf.\ also figs.\ 4, 5) that the approximation based on the
exactly solvable free fermion model yields a numerical value
of the critical hopping parameter fairly close to the
results of previous computer studies \cite{mus1},
\cite{gaus3}, \cite{karsch}. As mentioned above,
systematic improvements can be obtained by taking
into account correlation functions of $C$. This apparently
is necessary as one learns from figs.\ 4, 5 if one wants to
find the critical line beyond the region defined by the
critical points of the ordinary loop model and
the strong coupling Schwinger model respectively.\\

\section{Discussion and conclusions}

\parindent1.5em

Let us first have a look at the larger picture emerging for the critical
behaviour of the self-avoiding loop model with variable bending rigidity.
There is one point on the critical line known exactly from the
solution of the free fermion model ($\eta = 1/\sqrt{2}$, $z_{cr} = 2$)
\cite{pri1}-\cite{blu}. For this model it is established that the phase
transition is Ising-like, i.e., the model experiences a second order phase
transition with exactly the same critical exponents as the regular
Ising model. By the argument of universality we may conclude
that neighbouring models which lie on the same critical line
exhibit the same behaviour. This in particular concerns the
ordinary loop model ($\eta = 1$) and the strong coupling Schwinger
model ($\eta = 1/2$). For the ordinary loop model this has been
confirmed by MC investigations in the past \cite{kar3}.
For the strong coupling Schwinger model this consideration
specifies the so far unknown character of the phase
transition and confirms the recent suggestion that the transition
might be a continuous one \cite{karsch}.\\

In order to extend the understanding of the self-avoiding
loop model with variable bending rigidity at criticality
let us consider the central charge $c$ of the corresponding
conformal field theory (CFT). Helpful information can be
obtained most easily for the free fermion model considered
in sect.\ 4. First, it seems worthwhile mentioning that
the regular Ising model can be understood as a special free fermion
model \cite{sam}. A preliminary investigation along the lines given
for the Ising model in \cite{itz} indicates that the self-avoiding
loop model with bending rigidity $\eta = 1/\sqrt{2}$
can be represented at the critical point $z_{cr} = 2$ by just one
massless (continuum) Majorana fermion (As in the special case of
the Ising model just one half of the fermionic modes needed to
express the partition function becomes massless at the
critical point.). Consequently, this suggests that the critical
self-avoiding loop model at the free fermion point is equivalent
to a $c = 1/2$ CFT. The central charge cannot change continuously on
the critical line in the neighbourhood of the free fermion model,
therefore CFT's corresponding to the self-avoiding loop
model with variable bending rigidity should all be expected
to exhibit $c=1/2$. This of course entails that the strong
coupling Schwinger model at criticality should be equivalent to a
$c=1/2$ CFT. Consequently, in accordance with Zamolodchikov's c-theorem
\cite{zam1}, \cite{zam2},  $z \neq 0$ is related to a flow
 from the 6-vertex model ($z = 0$) having $c=1$ \cite{kar4},
\cite{kar5} towards a model with $c=1/2$
(as discussed in general terms by Salmhofer \cite{salm}).\\

We are going to test the above insight now by calculating the
central charge for the strong coupling Schwinger model.
This can be done most easily by considering the model
on a strip of width $a$ and length $b\rightarrow\infty$
\cite{blo}, \cite{aff}. The central charge is related to
the partition function (on a torus) by the formula
($f$ is the (bulk) free energy density on the infinite plane)

\parindent0.em

\begin{eqnarray}
\label{E1}
\lim_{b\rightarrow\infty}\
{\ln Z_{\Lambda(a\times b)}\left[z_{cr},1/2\right]\over b}&=&
af(z_{cr},1/2)\ +\ c\ {\pi\over 6}\ {1\over a}\ \ \ \ .
\end{eqnarray}

We however will approach the study of the central
charge of the strong coupling Schwinger model
by means of the exact partition functions calculated
on finite lattices in \cite{gaus3}, \cite{karsch}.
For our purpose the exact partition functions
available on a $8\times 8$ torus \cite{gaus3}, and on
$2\times 32$, $3\times 16$ tori \cite{karsch} are suited.
In fig.\ 6 we have plotted the function

\begin{eqnarray}
\label{E2}
c(z)&=&{a\over b}\ {6\over\pi}\
\left\{\ln Z_{\Lambda(a\times b)}\left[z,1/2\right]\ -\
ab\ f(z,1/2)\right\}
\end{eqnarray}

for the $2\times 32$, $3\times 16$ tori in dependence
on $\kappa$ ($z=(\kappa)^{-2}$ has been inserted).
$f(z,1/2)$ has been calculated by means of the
$8\times 8$ partition function. For sufficiently large
$b$ the function $c(z)$ should be expected to approach the value of the
central charge at the critical point. However,
one has to be aware of the fact that on the very narrow (with
respect to $a$) tori considered massless and massive fields
can contribute comparable amounts to the Casimir energy.
Inasmuch as the central charge is calculated by means of eqs.\
(\ref{E1}), (\ref{E2}) from the Casimir energy results obtained
 from very narrow tori may turn out misleading. In part, this is
what we observe from fig.\ 6. The result for the $a=2$ torus (solid line)
rather suggests $c = 1$ (or some value close to it),
however the torus is so narrow that
massless and massive fields contribute comparably to the Casimir energy.
Consequently, in agreement with our expectation $c = 1/2$
for the wider $a = 3$ torus (dashed line) we already observe a
much smaller value of $c(z)$ at the critical point. However,
it turns out that the sizes of the tori for which the exact
partition functions have been calculated so far in the literature
are too small to allow any final conclusions for the central
charge of the strong coupling Schwinger
model at criticality. Only partition functions calculated on
considerably larger lattices will allow to numerically test
the prediction $c=1/2$ in a reliable way.\\

\parindent1.5em
To conclude, the study of the self-avoiding loop model with
variable bending rigidity presented in this article
enhances the understanding
of the critical behaviour of the strong coupling Schwinger
model with Wilson fermions. We find that a second order
phase transition which lies in the Ising model universality
class takes place at some finite value of the hopping
parameter $\kappa_{cr}(0)$. Using certain approximate
analytic methods the value of the critical
hopping parameter is confirmed to lie at
$\kappa_{cr} \simeq 0.38-0.39$ in accordance with
earlier numerical investigations \cite{gaus3}, \cite{karsch}.
Certain arguments considered suggest that the strong
coupling Schwinger model at criticality is equivalent to
a $c=1/2$ CFT.\\

 From a technical point of view, the present paper
studies the application of the independent loop
approximation to the qualitative and in part quantitative
exploration of the phase structure of the self-avoiding
loop model with variable bending rigidity in 2D. Comparison
with known numerical results \cite{mus1} shows that
this method delivers a fairly correct picture for
sufficiently low (polymer) loop densities. This is
encouraging because the method is equally applicable to
higher dimensions, while the analytic approach based
on the exactly solvable free fermion model presented in
section 4 is at least in part specific to 2D. This suggests that the
independent loop approximation might successfully be
applied also to analogous systems in higher dimensions
(e.g., to strong coupling QCD in 4D where the critical
hopping parameter has recently been studied by other
methods \cite{gall}).\\

\newpage
\noindent
{\bf Acknowledgements}\\

The present work has been performed under the EC Human
Capital and Mobility Program, contract no.\ ECBCHBGCT930470.
I would like to thank Simon Hands for discussions and
helpful comments on a draft version of the paper.

\newpage

\section*{Figure captions}

\parindent.0em

{\bf Fig.\ 1}

Vertices of the 8-vertex model and their weights (cf.\ eqs.\
(\ref{B4a})-(\ref{B4d})).\\

{\bf Fig.\ 2}

Critical line of the self-avoiding loop model with variable
bending rigidity as found by MC calculations on a $64 \times 64$
lattice by M\"user and Rys \cite{mus1}. This figure has
been copied and adapted from fig.\ 2 of \cite{mus1}.
The dashed line indicates some interpolated curve
to the point $s = 1$, $T = 0$ (specially emphasized by
a black half disk). The domains
$\tt I-VI$ are mapped to the correspondingly labelled domains in
the $\{z,\eta\}$-plane (see fig.\ 3, boundary lines are plotted in the
same style in both figures). $\tt ff$ denotes the exactly
known critical point $\{T_{cr} = 2/(3\ln 2) \approx 0.962,\ s_{cr} = 1/3\}$
($z_{cr} = 2$) of the free fermion model ($\eta = 1/\sqrt{2}$)
\cite{pri1}-\cite{blu}. $\tt lg$ stands for the ordinary
self-avoiding loop model ($\eta = 1$, $s = 0$) with the
critical point $T_{cr} = 1.157$ ($z_{cr} = 2.373$)
\cite{barb}--\cite{kar3}. $\tt sm$ denotes the critical point
$\kappa_{cr}(0) = 0.38$ ($T_{cr} = 0.80$, $s_{cr} = 0.56$,
$z_{cr} = 1.73$) of the strong coupling Schwinger model
($\eta = 1/2$) as found in \cite{gaus3}, \cite{karsch}.\\

{\bf Fig.\ 3}

This is the equivalent of fig.\ 2 shown here for the
$\{z,\eta\}$ coordinate system.
For further explanations refer to fig.\ 2.\\

{\bf Fig.\ 4}

The critical line according to the results of the
independent loop approximation (\ref{C20}) (dotted line)
and of the free fermion model related approach (\ref{D13})
(dashed line) in comparison with the MC result (solid line) of
M\"user and Rys \cite{mus1}. For further explanations
refer to fig.\ 2.\\

{\bf Fig.\ 5}

This is the equivalent of fig.\ 4 shown here for
the $\{z,\eta\}$ coordinate system and relating to eqs.\
(\ref{C19}) (dotted line) and (\ref{D12}) (dashed line).\\

{\bf Fig.\ 6}

The function $c$ (see eq.\ (\ref{E2})) in dependence on $\kappa(0)$.
The solid line is the result for the $2\times 32$ lattice
while the dashed line stands for the result on the
$3\times 16$ lattice. The value of $c$ at the critical point
$\kappa_{cr}(0) \simeq 0.38-0.39$ has to be compared with
the expectation for the central charge (For a discussion see
the main text.)\\

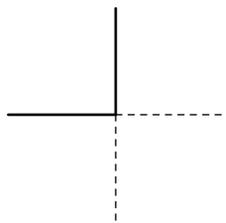
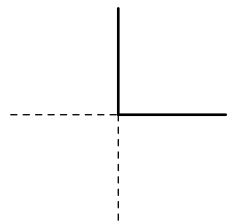
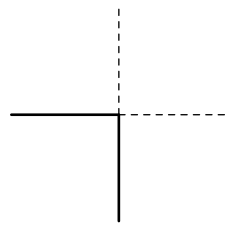
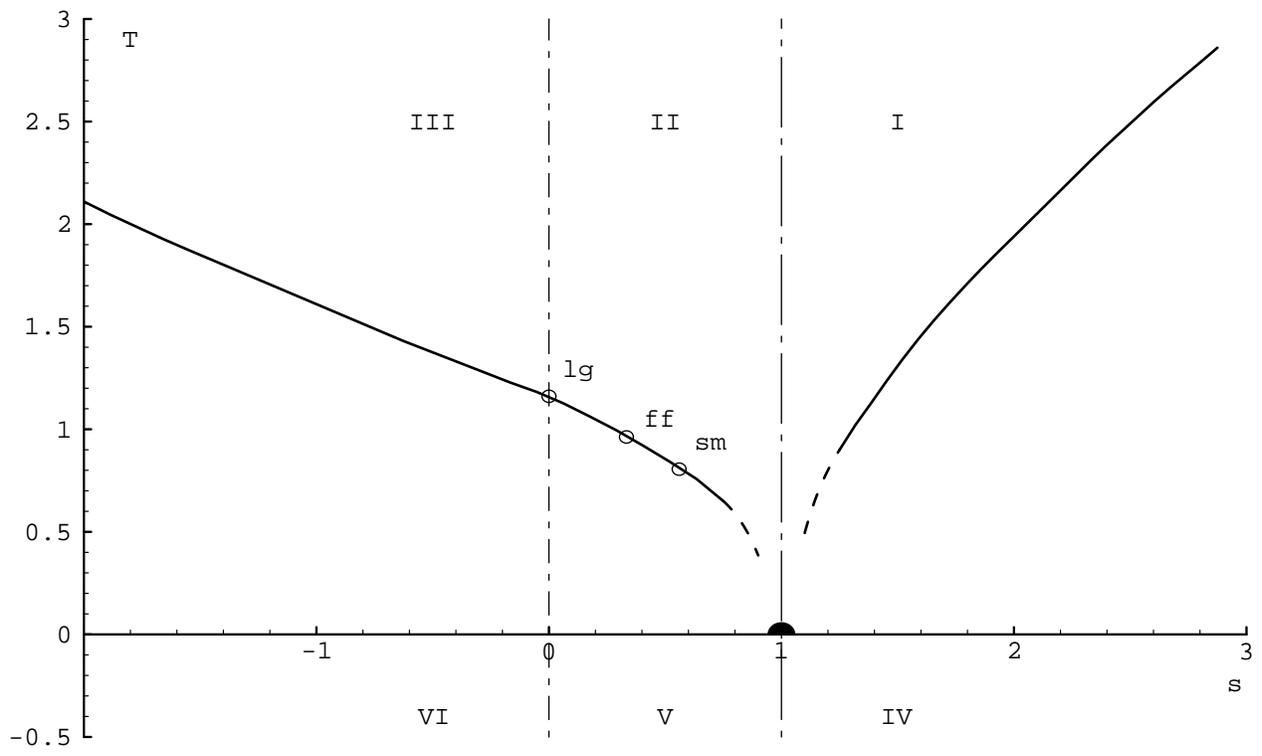
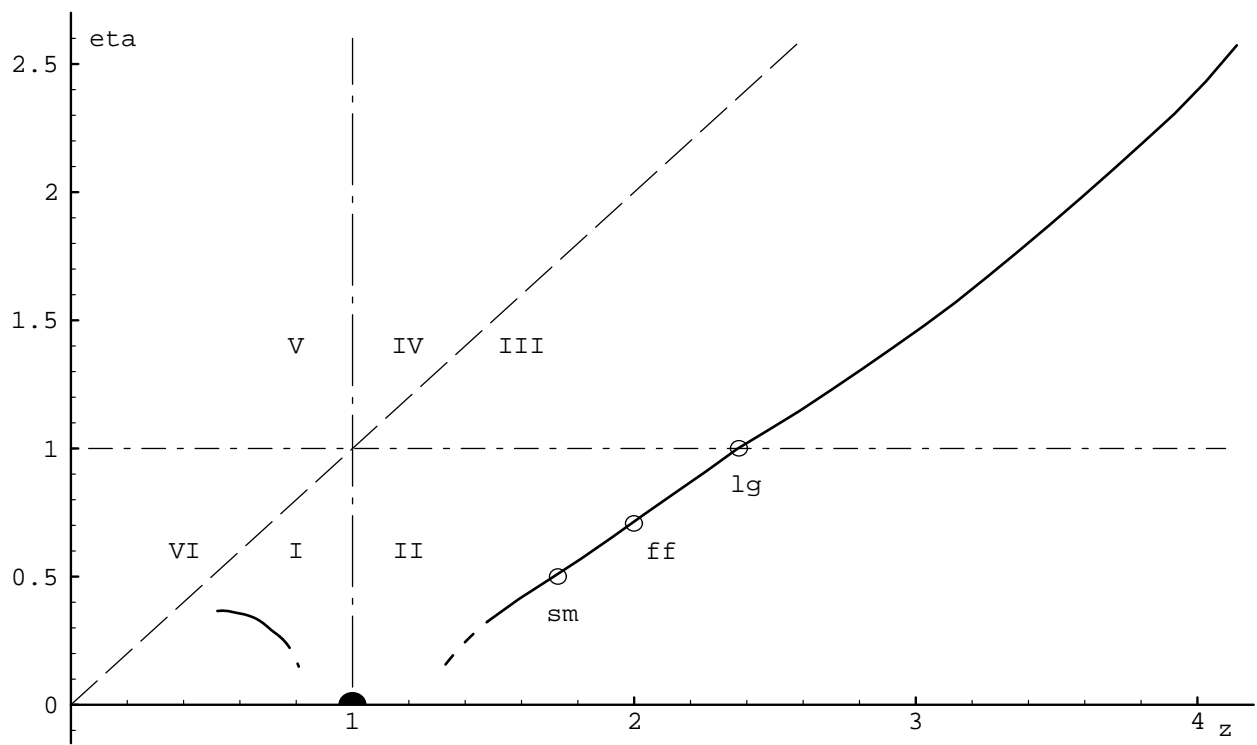
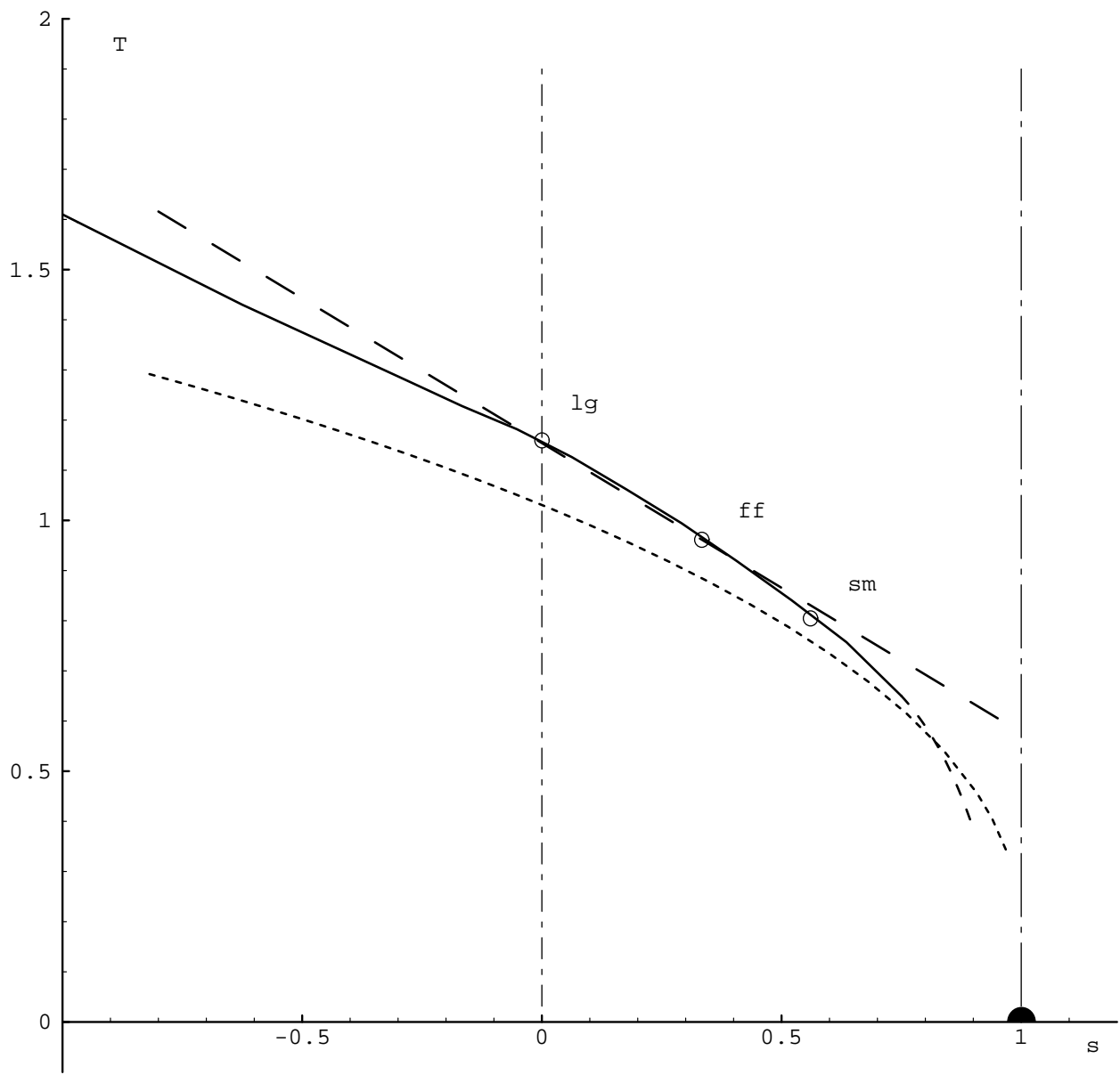
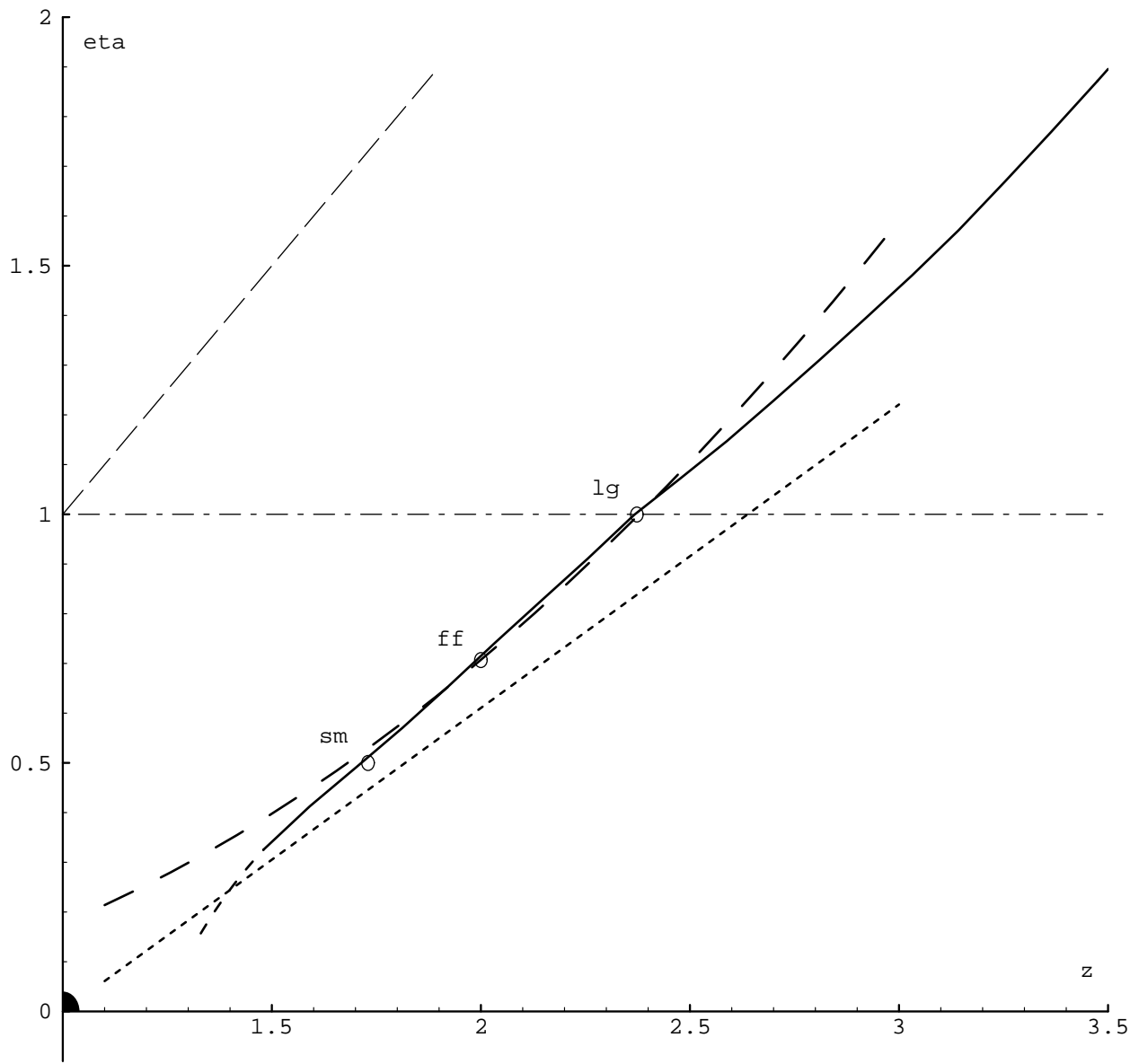
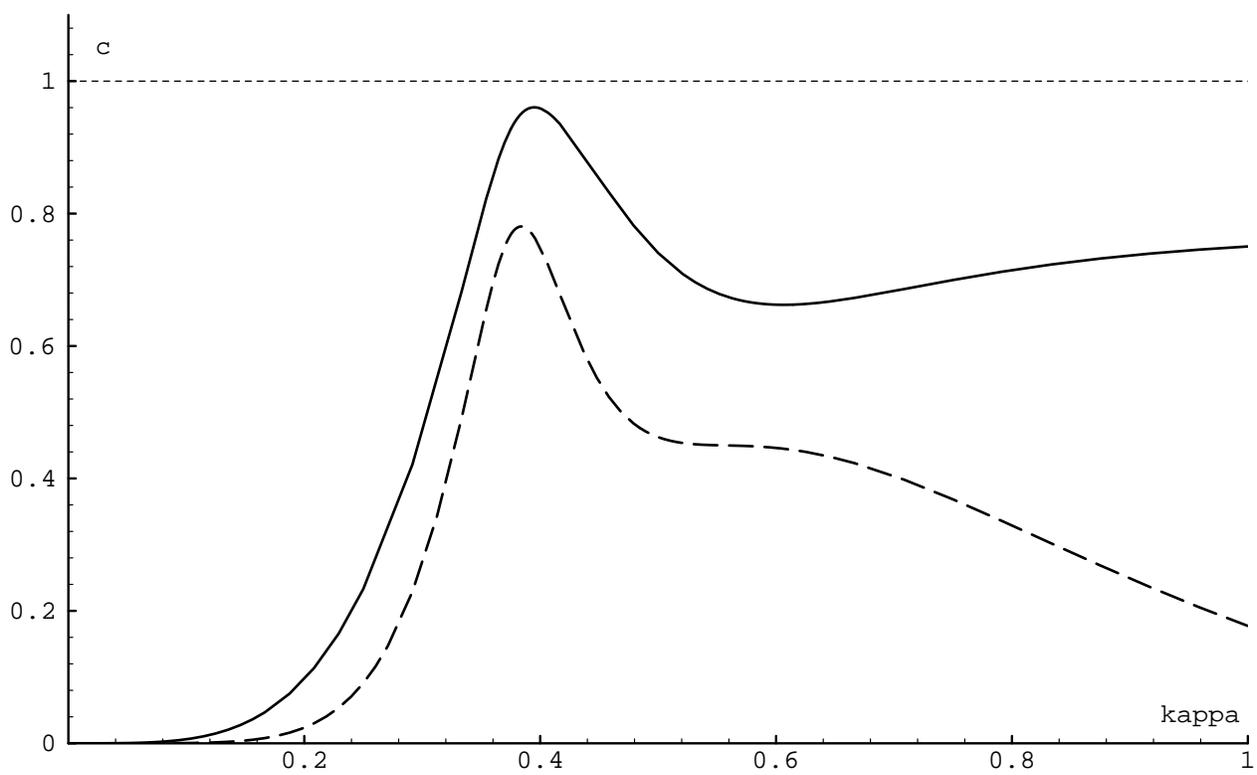
\end{document}